\documentclass[paper]{JHEP3}

\usepackage{amsfonts}
\usepackage[centertags]{amsmath}
\usepackage{amssymb}
\usepackage{epsfig}
\usepackage{subfigure}
\usepackage[section]{placeins}
\providecommand{\mivector}[1]{\vec #1 \,}
\providecommand{\bra}[1]{\langle #1\rvert}
\providecommand{\ket}[1]{\lvert#1 \rangle}

\providecommand{\proarrow}[0]{\rightarrow}

\providecommand{\dif}[0]{\mathrm{d}}

\providecommand{\proname}[2]{#1 \proarrow #2}
\providecommand{\lrproname}[2]{#1 \leftrightarrow #2}

\providecommand{\abs}[1]{\left\lvert #1 \right\rvert}
\providecommand{\abst}[1]{\bigl\lvert #1 \bigr\rvert}

\author{
Juan Racker$\,^{a, b}$    
and Esteban Roulet$^a$ \\
$^a${
\it CONICET, Centro At\'omico Bariloche, Avenida Bustillo 9500 (8400)
Argentina.}\\ 
$^b${
\it Departament d'Estructura i Constituents de la Mat\`eria and 
ICCUB, Institut de Ci\`encies del Cosmos,  Universitat
  de Barcelona, Diagonal 647, 08028 Barcelona, Spain.}\\
E-mail: {\rm \tt racker@ecm.ub.es, roulet@cab.cnea.gov.ar}}

\title{Leptogenesis, $Z'$ bosons, and the reheating temperature
  of the Universe}

\abstract{
We study the impact for leptogenesis  of new U(1) gauge  bosons coupled to the
heavy Majorana neutrinos. They can significantly enhance the efficiency of
thermal scenarios in the weak washout regime 
 as long as the $Z'$ masses are not much larger than the
reheating temperature ($M_{Z'}<10 T_{rh}$), with the highest efficiencies 
obtained for $Z'$ bosons considerably heavier than the heavy neutrinos
($M_{Z'} \gtrsim 100 M_1$). 
We show how the allowed region of the parameter space is modified in the
presence of a $Z'$ and we also obtain the minimum reheating temperature that
is required for these models to be successful.   
}

\begin{document}

\section{Introduction}
Leptogenesis is one of the most attractive known theories to explain the
origin of the matter-antimatter asymmetry of the universe~\cite{fukugita86}. This is because
it's based on a simple extension of the standard model (SM) which
can also explain naturally why the neutrino masses are so tiny. 
In leptogenesis scenarios there are two well different regimes according to the
strength of the Yukawa interactions, which is parametrized by the effective
mass $\tilde m_1$\footnote{\samepage{We will focus on hierarchical
  scenarios, for which the $N_1$ mass $M_1$ is much smaller than the masses of
  the other two heavy neutrinos $N_{2,3}$, i.e. $M_1\ll M_{2,3}$ and we also
  assume that the lepton asymmetry generated during $N_{2,3}$ decays is not
  relevant. The effective mass $\tilde m_1$ is just the $N_1$ decay width normalized to $M_1^2/8\pi v^2$.}}. The strong washout regime ($\tilde m_1 \gg 10^{-3}$~eV) is characterized by small
departures from equilibrium and a significant erasure of the asymmetries
generated by the heavy neutrino decays. On the other hand, in the weak
washout regime ($\tilde m_1 \ll 10^{-3}$~eV) the neutrinos decay far out from
equilibrium and the erasure of the asymmetry produced in the decay epoch is
negligible. Although the observed values of the differences of the squared
masses of the light neutrinos may suggest a high value for the effective mass, the
weak washout regime is well consistent with observations. In fact,
 the only bound
on $\tilde m_1$ coming from the light neutrino masses is $\tilde m_1 \ge m_1
\ge 0$. Nevertheless, the generation of a lepton asymmetry in this regime
faces some problems. 
If the SM is minimally extended adding only the heavy Majorana
neutrinos and one 
considers thermal leptogenesis, so that the heavy neutrinos are
produced by inverse decays and scatterings in the thermal bath, the production
of an asymmetry is limited by the small rate of production of the lightest
heavy neutrino.
In the traditional computation in which one includes scattering processes in
the production of $N_1$ but only considers the CP violation related to the
decays, inverse decays and $s$-channel off-shell scatterings, the
final baryon asymmetry turns out to 
be approximately proportional to $\tilde m_1$, being hence 
strongly suppressed for very small $\tilde m_1$. On the other hand, the
CP violating asymmetry per decay, $\epsilon$, can become larger for increasing
heavy neutrino masses~\cite{davidson02} 
and therefore thermal scenarios in the
weak washout limit ($\tilde m_1 \ll 10^{-3}$~eV) require large values of $M_1$
to be successful. This is represented by the dotted curve in
fig.~\ref{fig:1} which delimits the region of the $\tilde m_1 - M_1$ space
allowed by observations ($n_B/s = (8.82 \pm 0.23) \times
10^{-11}$~\cite{komatsu08})\footnote{Since we are interested in the weak washout regime we take
$m_1 \approx 0$, so that the bound~\cite{davidson02,hambye03} on the CP
asymmetry in $N_1$ decays becomes: $\abs{\epsilon} \leq
\tfrac{3}{16 \pi} \tfrac{M_1}{v^2} m_3$, with $m_3 \simeq \sqrt{\Delta
  m_{32}^2} \simeq 0.05$~eV.}. But the situation is actually 
worse than usually stated because there's a cancellation between the asymmetry
generated at early times during the production of $N_1$ and the asymmetry of
opposite sign generated during the decays. This cancellation shows up only
with the proper inclusion of
the CP asymmetries in scatterings~\cite{abada06II}, \cite{nardi07II} and a
final symmetric universe can be avoided only thanks to the action of the early
washouts which erase some of the ``wrong sign'' asymmetry produced in the
first stages. The weaker the 
early washouts are, the less asymmetry survives this cancellation, and taking
this into account the final baryon
asymmetry is actually proportional to approximately $\tilde m_1^2$ and not just to
$\tilde m_1$. The resulting lower bound on $M_1$ is represented by the solid
curve of fig.~\ref{fig:1}, which in the weak washout regime is clearly stricter than the traditionally
quoted bound.
\FIGURE{
\centerline{\protect\hbox{
\epsfig{file=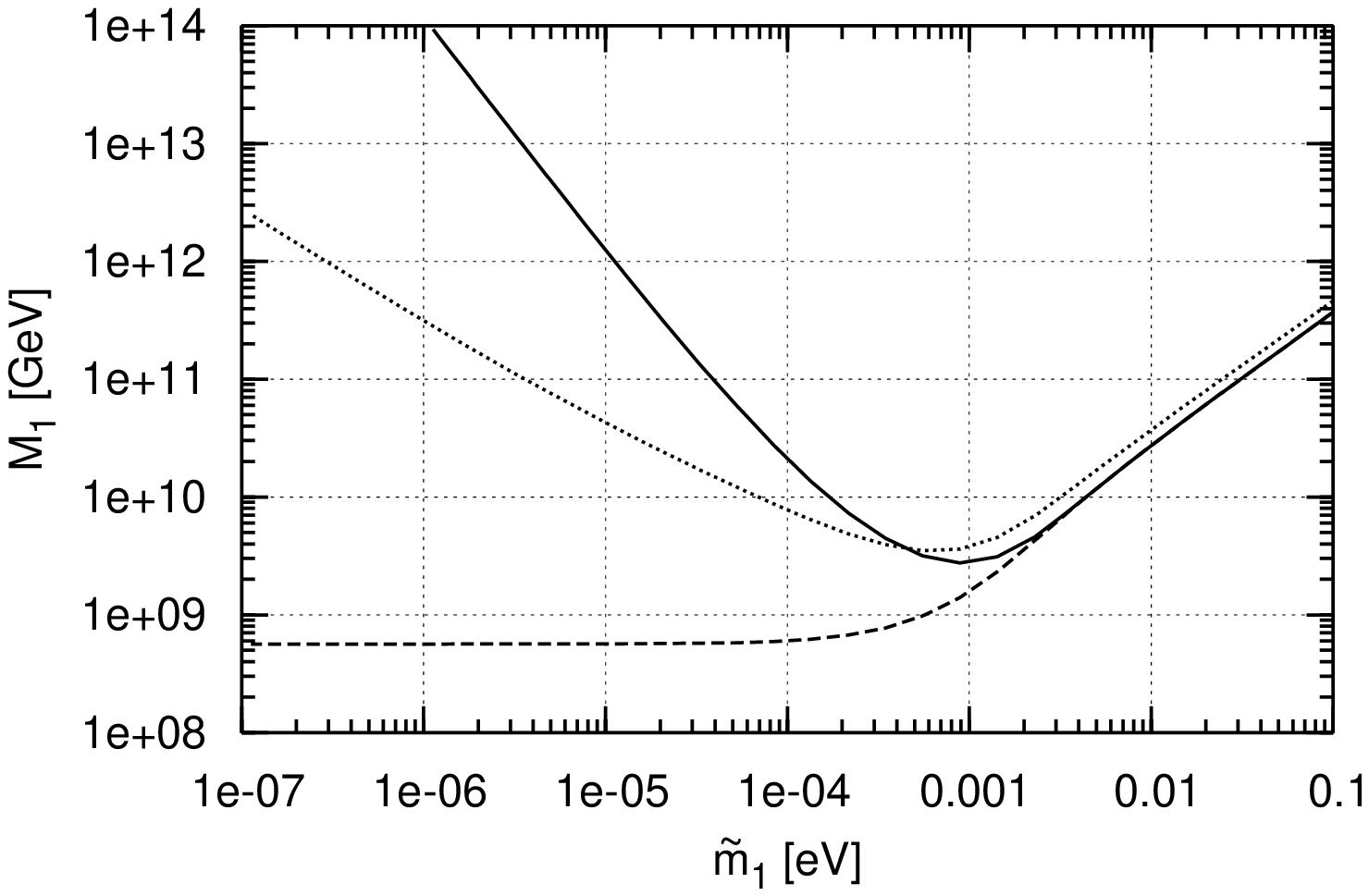 ,width=0.5\textheight,angle=270}}}
\caption[]{Allowed regions in the $\tilde m_1 - M_1$ plane assuming a null
  initial $N_1$ density, not including the CP violation in scatterings (region
  above the dotted line) and including it (region above the solid line). The
  allowed region for the case of an equilibrium initial density is also
  represented (region above the dashed line). The lines are the values leading
  to $Y_B=8.7\times 10^{-11}$ adopting the CP asymmetries saturating the Davidson-Ibarra bound~\cite{davidson02}.} 
\label{fig:1}
}  

The high Majorana neutrino masses required to produce the observed asymmetry
in the weak  
washout regime may be in conflict with the relatively low reheating temperature
of many cosmological models and this is also a potential problem in
supersymmetric scenarios affected by the possible overproduction of gravitinos. 

The situation just described is quite different
 if the abundance of the
heavy neutrinos at the beginning of the leptogenesis era is, for some reason,
equal to that of equilibrium. In this case the final
$B-L$ asymmetry in the weak washout regime is  $Y_{B-L} \approx - \epsilon
 Y_N^{eq} (T \gg M_1)$ (where $Y_i\equiv n_i/s$ and $N \equiv N_1$), 
i.e. the efficiency\footnote{The efficiency $\eta$ is defined by $Y_{B-L}^f = -
 \epsilon \eta  Y_N^{eq} (T \gg M_1)$, where $Y_{B-L}^f$ is the final $B-L$ asymmetry.} is approximately
equal to unity and the allowed region in the $\tilde m_1 - M_1$ plane is
greatly 
enlarged (dashed line in fig.~\ref{fig:1}). One of the ways to reach an
equilibrium density before the onset of leptogenesis is to have new
interactions. In particular, the case for gauge interactions was considered
in~\cite{plumacher96}, \cite{cosme04}. Here we are going to extend and study
in more detail the scenarios with $Z'$ bosons. In section~\ref{sec:model}
we define the model we are going to concentrate on, we then describe 
in section~\ref{sec:effects} the different 
 effects induced by the presence of the $Z'$  and in
section~\ref{sec:RT} we study situations in which the reheating
temperature after inflation is less than the mass of $N_1$ and determine the
 minimum reheating temperature compatible with successful leptogenesis. The
conclusions are presented in section~\ref{sec:concl}.     

\section{The model and Boltzmann equations}
\label{sec:model}
In order to give numerical results we are going to work with a specific
model, but the features
that will be described are expected to be valid for different models that include
a new neutral gauge boson. We will use the model described in~\cite{plumacher96} and here we give a brief summary of it, emphasizing the
most relevant points for our work. 

The gauge symmetry of the SM is extended to the group $SU(3)_C \times
SU(2)_L \times U(1)_Y \times U(1)_{Y'}$, which can arise as a step in the
chain of spontaneous symmetry breakings from an unification gauge group like
$SO(10)$ to the SM one. The covariant
derivative is 
\begin{equation*}
D_\mu = \partial_\mu -ig\mivector{W}_\mu \cdot \mivector{T} -ig'B_\mu Y -
ig'\sqrt{\frac{2}{3}} C_\mu Y'\; ,
\end{equation*} 
where $\mivector{W}_\mu, B_\mu$ and $C_\mu$ are the $SU(2)_L, U(1)_Y$ and $
U(1)_{Y'}$ gauge fields respectively. Note that both abelian groups have the
same gauge coupling constant ($g'$) as a consequence of their (assumed) common
origin in the larger group $SO(10)$. It can also be shown that the $U(1)_{Y}$
and $U(1)_{Y'}$ charges of a particle are related by $Y' = Y -
\tfrac{5}{4} (B-L)$, which gives the coupling between the different fermions
of the model and the $U(1)_{Y'}$ gauge field in terms of the (known) weak hypercharges. 

The scalar sector of the model consists of the SM Higgs and the field
$\chi$ responsible for the spontaneous symmetry breaking (SSB) of $U(1)_{Y'}$
at a scale $v' = \bra{0}\chi\ket{0}  \gg  v$ (with $v=174$~GeV the vacuum
expectation value of the SM Higgs). Due to the SSB of $U(1)_{Y'}$ the right
handed neutrinos acquire a Majorana mass given by $M = y v'$, with
$y$ the matrix of Yukawa couplings between the right handed neutrinos and the
$\chi$ field. The $U(1)_{Y'}$ gauge field also becomes massive:
$M_{Z'}=\tfrac{5}{\sqrt{3}} g' v'$, where $Z'$ is the massive $U(1)_{Y'}$
gauge boson. We will assume that $M_{Z'}$ is larger than $M_1$, which is natural given that gauge
couplings are usually larger than the Yukawa ones. As explained in~\cite{plumacher96}, the Higgs boson associated to the SSB of
$U(1)_{Y'}$ can be neglected when studying processes that occur at
temperatures below $M_{Z'}$, in particular during the $N_1$ leptogenesis
era.   

The most relevant
processes for the thermalisation of the heavy neutrinos are those mediated by $Z'$ which produce or
destroy the heavy Majorana neutrinos, i.e. $\lrproname{f \bar f, h \bar h}{N_j
  N_j}$ (where $f$ is a SM fermion and $h$ is the SM Higgs). We have calculated the reduced cross section obtaining\footnote{The
  expression given in eq.~\eqref{eq:cs} differs from that obtained
  in~\cite{plumacher96} and also with that in~\cite{cosme04}.}:
\begin{equation}
\label{eq:cs}
\hat \sigma_{Z'} (s) = \frac{4225\pi}{216} \frac{\alpha^2}{\cos^4 \theta_w}
\sqrt{x({x - 4a_j})}  \frac{x-a_j}{(x-a_{Z'})^2 + a_{Z'}c}\; , 
\end{equation}
where all quantities with dimension of energy are normalized to the mass of
the lightest heavy Majorana neutrino: $x \equiv s/M_1^2, a_j \equiv (M_j/M_1)^2, a_{Z'} \equiv (M_{Z'}/M_1)^2$
and $c \equiv (\Gamma_{Z'}/M_1)^2$, with $\Gamma_{Z'}$ the decay width of $Z'$
which is given by~\cite{plumacher96} $\Gamma_{Z'} = \tfrac{\alpha}{\cos^2
  \theta_w} M_{Z'} \bigl[ \tfrac{169}{144} + \tfrac{25}{18} \sum_i \bigl(
\tfrac{a_{Z'} - 4a_i}{4a_{Z'}}\bigr)^{3/2}\theta(a_{Z'} - 4a_i) \bigr]$. The cross section given is summed over all
the degrees of freedom of the particles involved (the initial
particles considered are the SM Higgs and fermions). 

The processes $\lrproname{f \bar f, h \bar h}{N_j N_j}$ don't violate lepton
number, so they only enter in the Boltzmann equation for the evolution of
$Y_N$, which becomes\footnote{The processes involving the SM gauge bosons won't
be included in this work.}:
\begin{equation}
\frac{\dif Y_N}{\dif z}
= -\frac{1}{sHz}\left\{\left[\frac{Y_N}{Y_N^{eq}}-1\right]
     \left( \gamma_D+2\gamma_{Ss}+4\gamma_{St} \right) + \left[\left(
 \frac{Y_N}{Y_N^{eq}} \right)^2 - 1\right] \gamma_{Z'} \right\} \; ,
\end{equation}
with $z \equiv M_1/T$. The quantities $\gamma_D, \gamma_{Ss} (\gamma_{St})$ and $\gamma_{Z'}$ are the
reaction densities for decays, scatterings involving the top quark which are mediated by the Higgs in the $s$
($t$) channel and annihilation of $N_1$ pairs respectively. Note that, since
$\gamma_{Z'}$ doesn't depend on the Yukawa couplings of the heavy Majorana
neutrinos, the corresponding term in the Boltzmann equation has a different
dependence on the parameters of the model than the other three terms: while
these last are proportional to $\tilde m_1$, the $Z'$ term is inversely
proportional to $M_1$, so that for fixed values of $M_{Z'}$ and $\tilde
m_1$ the $Z'$ effects diminish for increasing values of $M_1$.

We end up this section with some comments about the conditions under which the Boltzmann
equations will be solved. Since we want to concentrate on the effects of the $Z'$ bosons we will consider
a simple flavor structure~\cite{abada06, nardi06}, assuming that the
$\tau$ is the only relevant lepton flavor. Anyway, flavor effects have a
limited impact in the weak washout regime which is the most relevant one for
this work. It's also necessary to specify the
fast processes that, while not entering directly in the Boltzmann equations
for $Y_N$ or $Y_{B-L}$, have influence on the generation of the
matter-antimatter asymmetry by redistributing the generated asymmetry among
the different particles of the thermal bath~\cite{buchmuller01, nardi05}. The set of spectator processes
that are active depends mainly on the value of $M_1$ since this determines the
typical temperatures of the leptogenesis epoch and to a smaller extent on $\tilde m_1$ because it establishes the duration of
this epoch. Nevertheless, we will always include the same set of spectator processes, namely that
corresponding to the temperature range $10^{11}~\rm{GeV} \lesssim T \lesssim
10^{12}~\rm{GeV}$~\cite{nardi05}, independently of the value of $M_1$ and
$\tilde m_1$, since these processes modify the final baryon asymmetry by only
some tens of percent, which is not important for our study of the $Z'$
effects. Finally, we will also ignore finite temperature corrections to the particle masses and
couplings~\cite{giudice04}. 
\section{The effects of $Z'$ in the weak washout regime}
\label{sec:effects}
The coupling of $N_1$ with the $Z'$ boson allows the production of the heavy
neutrinos without generating a CP asymmetry (contrary to the case of
production via the Yukawa couplings). This can help to solve the  
problems related with the production of the matter-antimatter asymmetry in the
weak washout regime, since the cancellation mentioned in the introduction may
no longer be enforced. But on the other hand, the
neutrinos can also be destroyed by these interactions without generating an
asymmetry, and this last effect  can reduce the efficiency of the production of a
cosmic asymmetry. 

\begin{figure}[!t]
\centering
\subfigure[]{\label{subfig:a}
\protect\hbox{
\epsfig{file=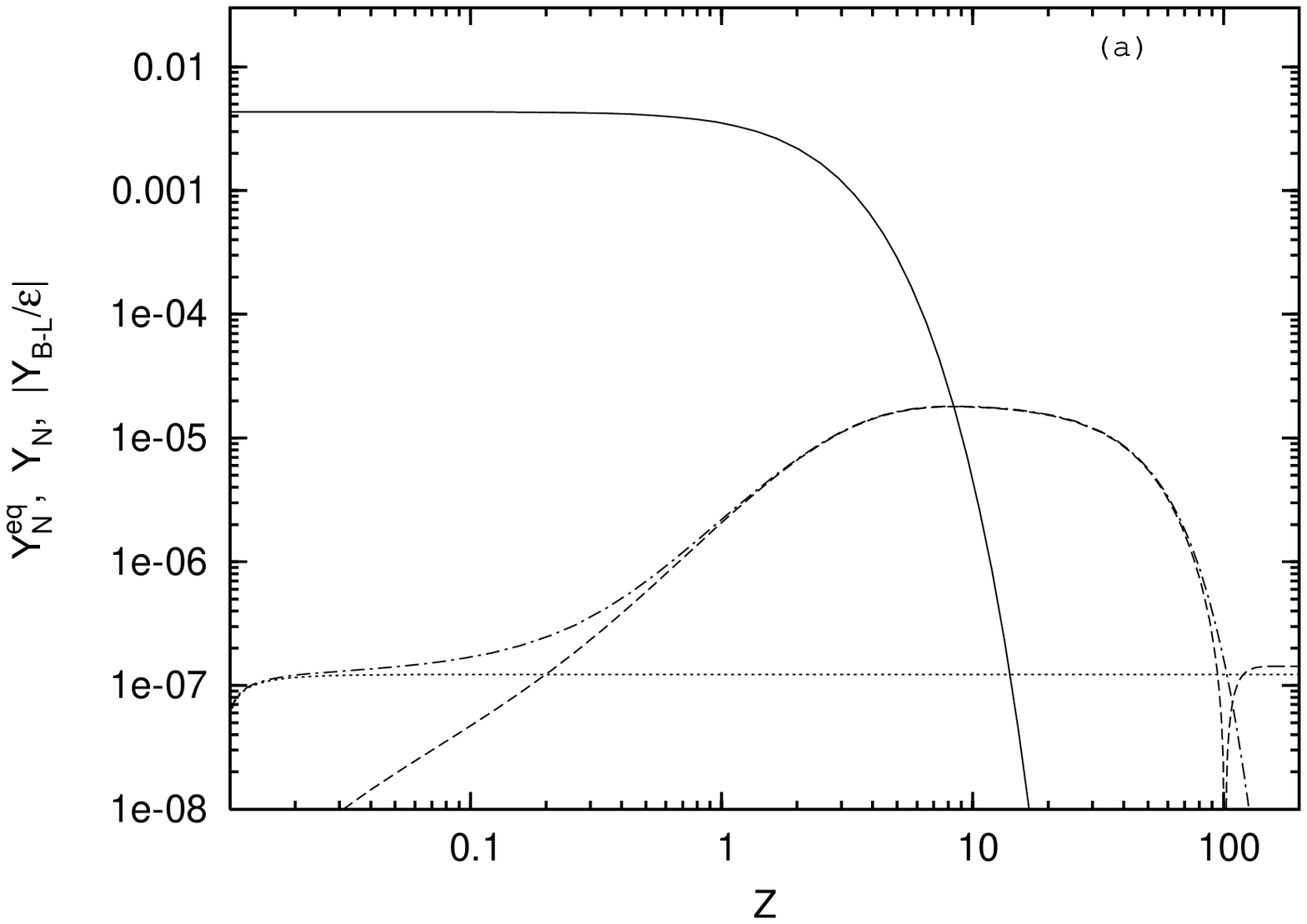,width=0.21\textheight,angle=270}}} $\quad$ 
\subfigure[]{\label{subfig:b}
\protect\hbox{
\epsfig{file=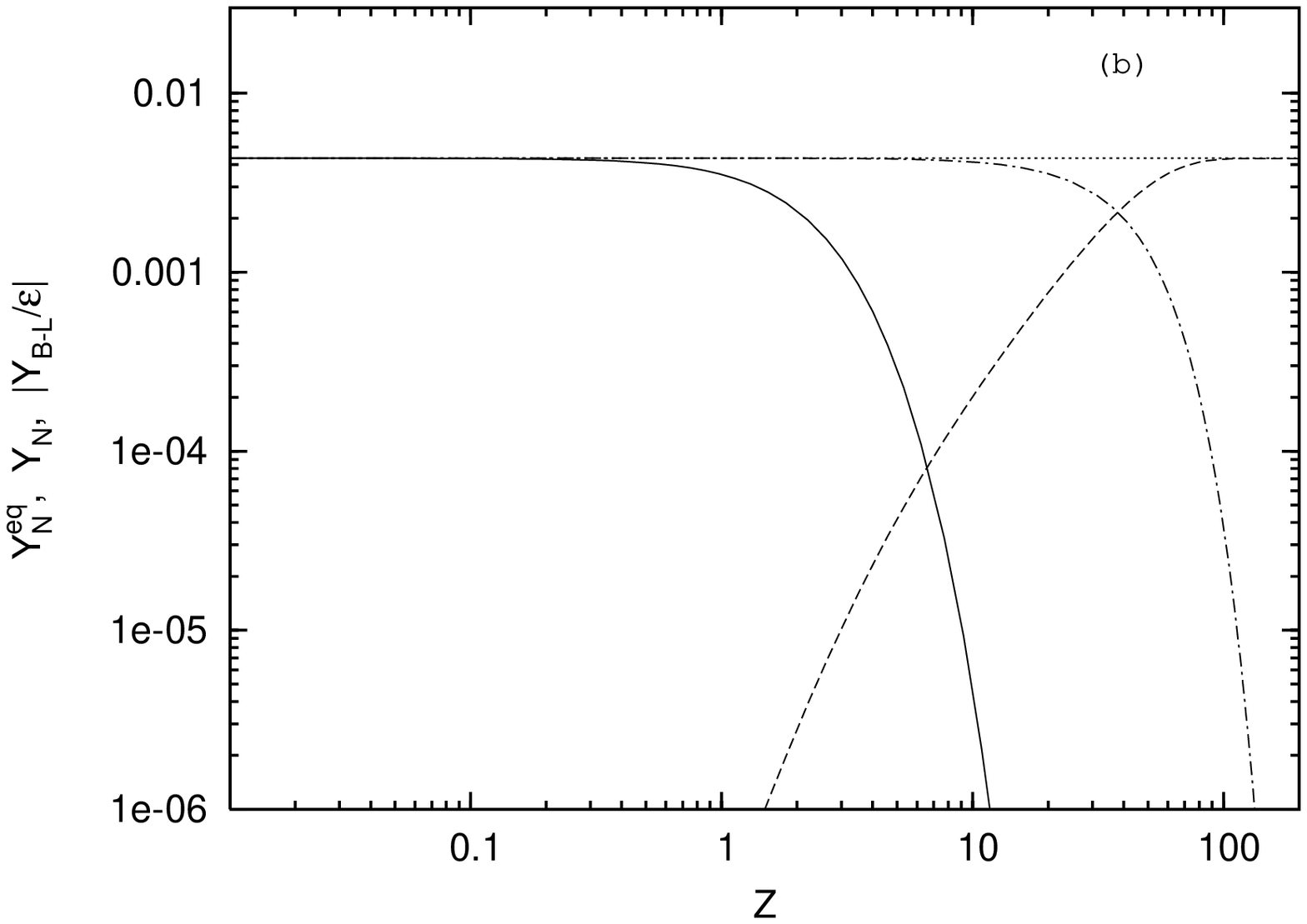,width=0.21\textheight,angle=270}}} \\
\subfigure[]{\label{subfig:c}
\protect\hbox{
\epsfig{file=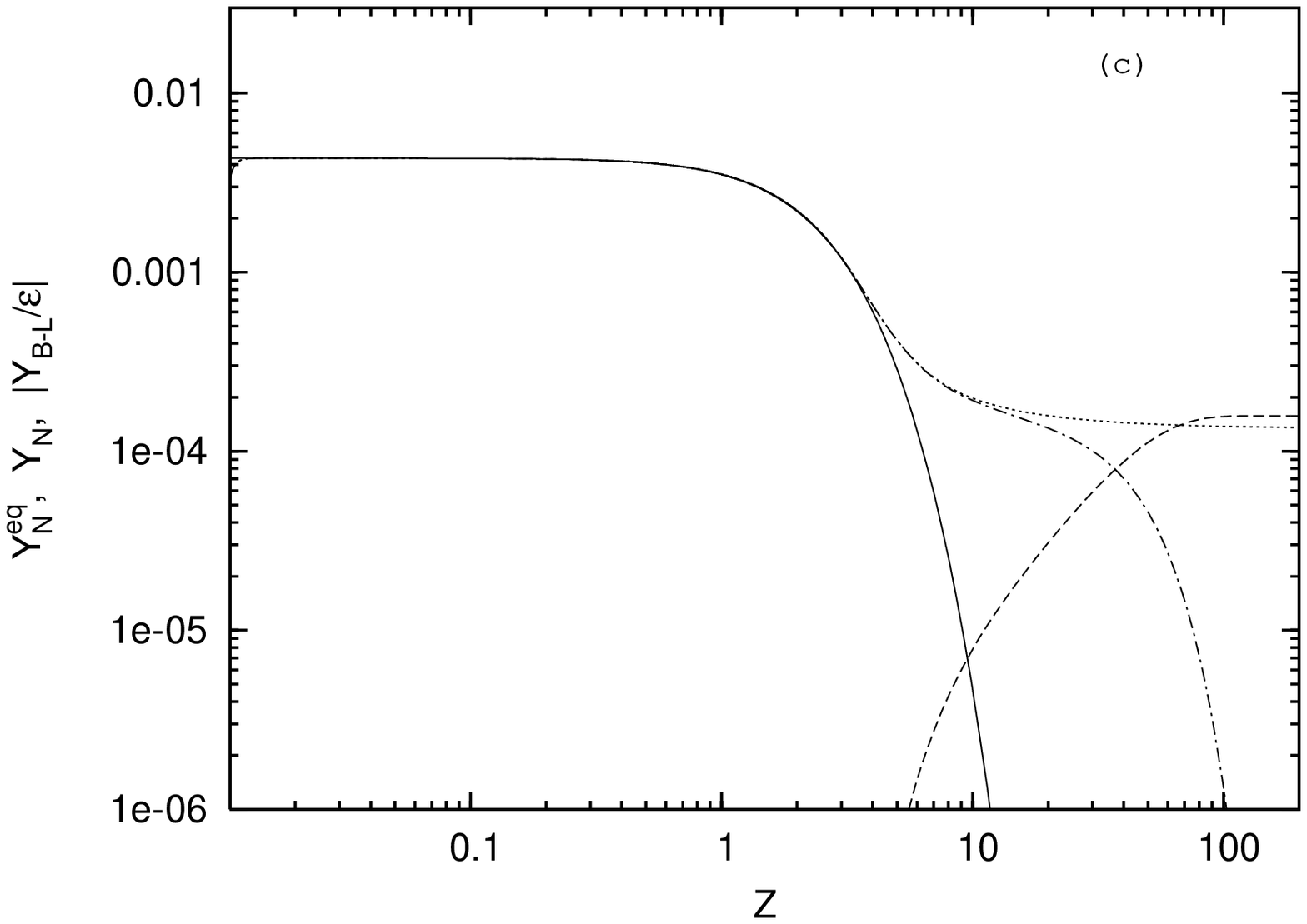,width=0.21\textheight,angle=270}}} $\quad$ 
\subfigure[]{\label{subfig:d}
\protect\hbox{
\epsfig{file=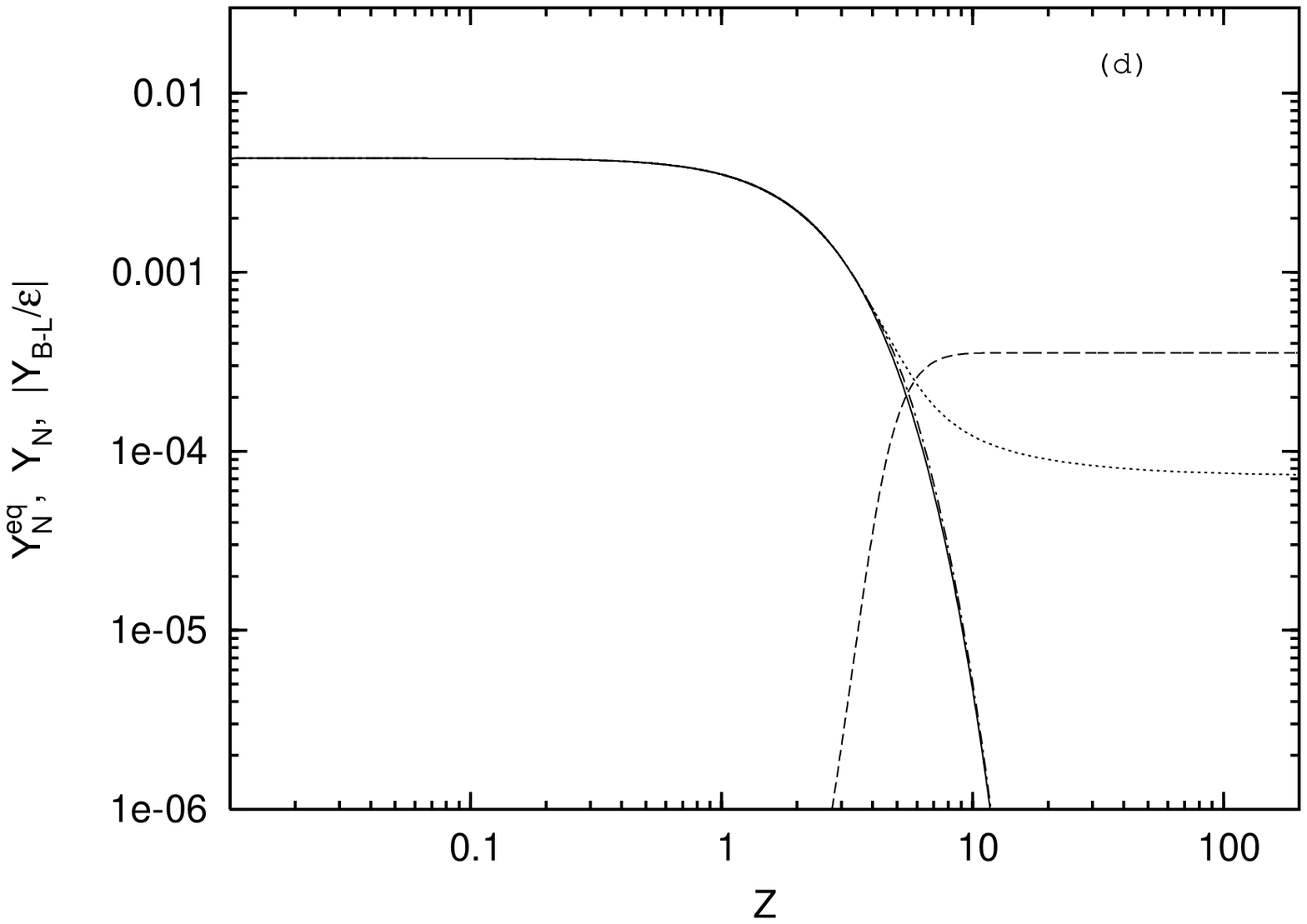,width=0.21\textheight,angle=270}}} 
\caption[]{The evolution of $Y_N^{eq}$ (solid line), $Y_N$ (dash-dotted
  line) and $\abst{Y_{B-L}/\epsilon}$ (dashed line) as a function of $z$ for different values of $M_{Z'}$. For comparison, the evolution of
  the  $N_1$ density assuming that the Yukawa interactions are null is also
  depicted (dotted curve). The values of $M_{Z'}, M_1$ and $\tilde m_1$ for
  each of the four panels (a), (b), (c) and (d) are those corresponding to the
  equally named points in fig.~\ref{fig:2}.}
\label{fig:3}
\end{figure}

Two important energy scales in the study of leptogenesis are the reheating
temperature
$T_{\text{rh}}$, which determines the initial time at which the heavy neutrinos
start to be produced in thermal scenarios, 
 and the heavy neutrino mass
$M_1$ which establishes the temperature at which the
equilibrium distribution of the heavy neutrinos starts to become Boltzmann suppressed. The effects of the
$Z'$ bosons depend on the value of its mass relative to these two scales.
To study quantitatively these effects let's first 
fix $T_{\text{rh}} = 100 M_1$, corresponding to a situation in which the thermal history of the universe starts well before the leptogenesis era. 
In fig.~\ref{fig:2} we depict the region in the $\tilde m_1 - M_1$
space that may lead to a sufficient generation of a baryon asymmetry 
 for different values of $M_{Z'}$.     
\FIGURE{
\centerline{\protect\hbox{
\epsfig{file=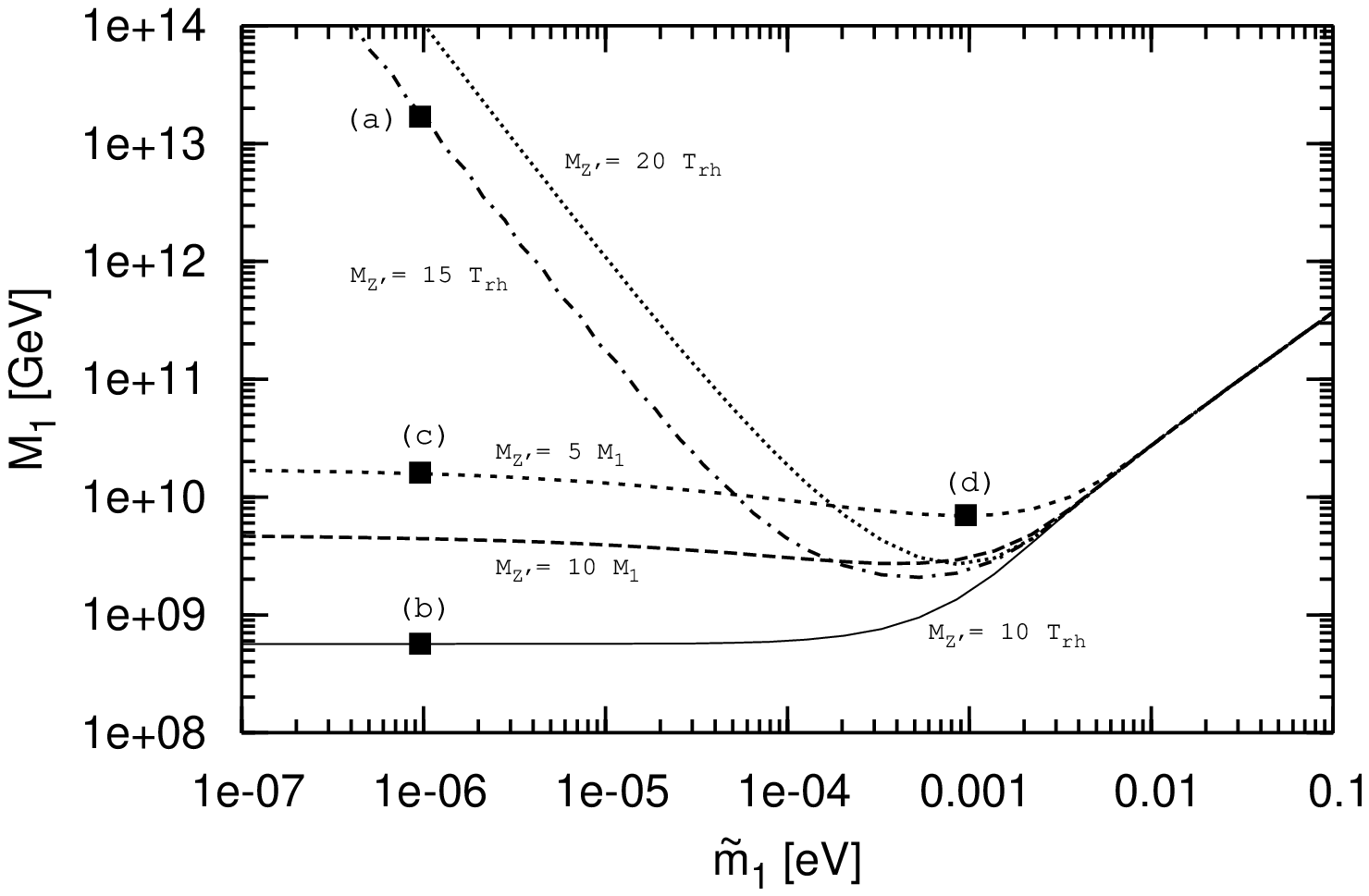 ,width=0.5\textheight,angle=270}}}
\caption[]{The regions allowed by observations in the $\tilde m_1 - M_1$
  parameter space for different values of the $Z'$ mass and $T_{rh}=100 M_1$. The regions allowed
  are those above the dotted line (for $M_{Z'} = 20 T_{rh}$), the
  dash-dotted line (for $M_{Z'} = 15 T_{rh}$), the solid line (for $M_{Z'} =
  10 T_{rh}$), the long dashed line (for $M_{Z'} = 10 M_1$) and the short dashed line
  (for $M_{Z'} = 5 M_1$). The points labeled (a) to (d) correspond to the
  panels of fig.~\ref{fig:3}, where the evolution of the $B-L$ asymmetry is represented.} 
\label{fig:2}
}  

Three different situations can be distinguished:
\begin{itemize}
\item[(i)] $M_{Z'} \gg 10 T_{\text{rh}}$: The $Z'$ boson is too heavy
  relative to the reheating temperature of the universe so that the associated
  cross section is very small and the effects of the $Z'$ are hence negligible
  (the curve in fig.~\ref{fig:2} corresponding to $M_{Z'} = 20 T_{rh}$ is similar
  to the solid line in fig.~\ref{fig:1}, which ignored the effects of new gauge bosons).
\item[(ii)] $100 M_1 \lesssim M_{Z'} \lesssim 10 T_{\text{rh}}$: An
  equilibrium population of $N_1$ is produced due to the new gauge interactions and
  these last depart from equilibrium before the $N_1$ become non-relativistic. This situation is optimal for the generation of a baryon
  asymmetry and the highest efficiencies are obtained. 

As can be seen in fig.~\ref{fig:2} the change from regime (i) to (ii) takes
place abruptly for $M_{Z'}$ in the range $(10-20) T_{rh}$ due to the fact that
for large $Z'$ masses $\hat \sigma_{Z'} \propto M_{Z'}^{-4}$. Although the
reheating temperature in fig.~\ref{fig:2} was fixed to $T_{rh} = 100 M_1$, a similar change is also found for other values of $T_{rh}$.
\item[(iii)] $M_{Z'} \lesssim 100 M_1$: The new gauge interactions are still in
  equilibrium when the heavy neutrinos become non-relativistic, so the $N_1$
  have a significant probability of disappearing without producing an
  asymmetry. For $M_{Z'} \gtrsim 3 M_1$ it's clear that the lighter the $Z'$ is, the
  later the gauge interactions fall out from equilibrium and the less
  asymmetry is then produced. However, for $Z'$ masses close to $M_1$ the
  analysis has to be done more carefully. The point is that, independently of
  the $Z'$ mass, this suppression
  effect is limited because the annihilation involves two heavy neutrinos and
  hence the corresponding reaction density is suppressed by two Boltzmann
  factors. On the other hand, when $M_{Z'}$ is close to $2 M_1$ the 
  reaction density is enhanced at $T \sim M_1$ because the $Z'$ that mediates
  the  annihilation can be produced resonantly, so the suppression
  effect induced by the $U(1)_{Y'}$ gauge interaction is maximum for $M_{Z'}
  \approx 2 M_1$\footnote{\samepage{If there were also charged gauge bosons
  $W_R^{\pm}$ associated to a right handed $SU(2)$ symmetry, the
  suppression effect on leptogenesis would be highly enhanced because of the existence of
  scatterings involving a single heavy neutrino and new $N_1$ decay channels
  which are CP symmetric~\cite{carlier99},\cite{frere08}.}}. It must also be
  noted that when the $Z'$ bosons are light, the effects of the Higgs field
  $\chi$ should also be taken into account.
\end{itemize} 
These results can be understood as the combination of two stages. In the first
one the gauge interactions dominate over the Yukawa interactions basically
until they depart from equilibrium at a temperature $T_{fo}$,
 leaving a relic density of $N_1$ which
will be similar to that of  a massless degree of freedom in equilibrium if $100
 M_1 \lesssim M_{Z'} \lesssim 10 T_{\text{rh}}$ while it will be Boltzmann
suppressed like the usual cold relics, with
density $ Y_N^{relic} \propto \exp(-M_1/T_{fo})$, if $M_{Z'} \lesssim 100
 M_1$. In the second stage, the neutrinos decay via their Yukawa
interactions producing a final asymmetry $Y_{B-L}^f \approx - \epsilon
Y_N^{relic}$.

As can be seen from fig.~\ref{fig:3} this picture explains the results very well
except when the $Z'$ is not very heavy and $\tilde m_1$ approaches the
equilibrium mass $m^*\simeq
10^{-3}$~eV (fig.~\ref{subfig:d}), since in this
case the Yukawa interactions begin to dominate over the gauge interactions
before these last depart from equilibrium. Note that in fig.~\ref{subfig:a},
which corresponds to a case in which  $M_{Z'} > 10 T_{\text{rh}}$ but still
the $Z'$ effects are important, the final
asymmetry is also given by $Y_{B-L}^f \approx - \epsilon
Y_N^{relic}$ but here $Y_N^{relic} \ll Y_N^{eq}(T \gg M_1)$ since the cross
section for pair production of $N_1$ mediated by the $Z'$ bosons is too small
after reheating to populate the universe with an equilibrium density of heavy neutrinos.  
\section{The reheating temperature}
\label{sec:RT}
The existence of $Z'$ bosons coupled to the heavy neutrinos also has
an impact on the lowest reheating temperature compatible with successful
leptogenesis. When $Z'$ bosons are absent, it has
been shown that the reheating temperature can be several times smaller than
$M_1$ in the strong washout regime~\cite{buchmuller04}. This
is due to the fact that the Yukawa interactions, being strong in this regime, can produce a
considerable amount of $N_1$ even if they begin to act when $T < M_1$. On the
other hand, the minimum reheating temperature for a given value of $\tilde m_1$
in the weak washout regime is approximately equal to the lower bound on $M_1$ 
for that value of $\tilde m_1$. 

The situation changes when the heavy neutrinos
can also be produced via gauge interactions. If the $Z'$ bosons are very
massive (cases labelled (i) and (ii) in the previous section), the gauge
interactions are already out of equilibrium at $T \sim M_1$ and therefore 
the reheating temperature has to be greater than $M_1$ in order to enhance the efficiency of leptogenesis by means of  the $Z'$ induced production of $N_1$
(see fig.~\ref{fig:4MZ100} for the case $M_{Z'}=100 M_1$);
but if
they are light (case (iii)) successful leptogenesis is possible for reheating
temperatures lower than $M_1$ also in the weak washout regime. This is shown
in fig.~\ref{fig:4MZ5} for $M_{Z'} = 5 M_1$, where the allowed regions of the
$\tilde m_1 - M_1$ plane for different values of $T_{rh}$ (relative to $M_1$) are
plotted. The allowed region is the same for all values of $T_{rh}$ satisfying
$T_{rh} \gtrsim M_1/3$, so the reheating temperature in this case can be up to
approximately three times smaller than $M_1$ for any value of $\tilde m_1$. On the other hand, for $T_{rh}
\lesssim M_1/5$ the allowed region is significantly reduced and doesn't
depend on the presence of the $Z'$ bosons (note that the curves corresponding
to $T_{rh} = M_1/5$ are almost the same in figs.~\ref{fig:4MZ100} and
\ref{fig:4MZ5}). 
\FIGURE{
\centerline{\protect\hbox{
\epsfig{file=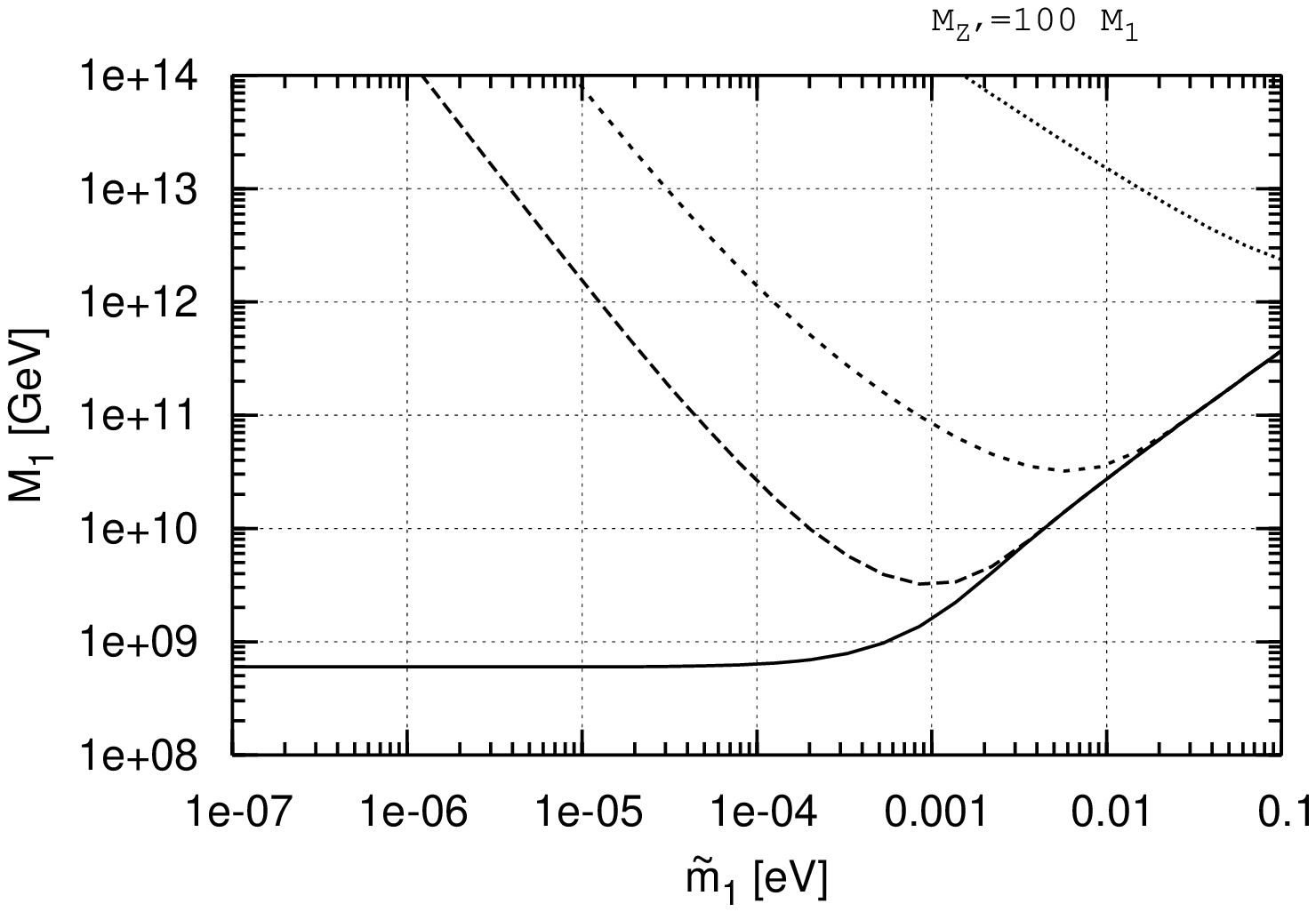 ,width=0.5\textheight,angle=270}}}
\caption[]{The allowed regions in the $\tilde m_1 - M_1$ plane for $M_{Z'}=
  100 M_1$ and different values of the reheating temperature: the regions
  above the solid, long dashed, short dashed and dotted curves
  correspond respectively to $T_{rh} = 10 M_1, M_1, M_1/5$ and $0.1 M_1$.} 
\label{fig:4MZ100}
}  

\FIGURE{
\centerline{\protect\hbox{
\epsfig{file=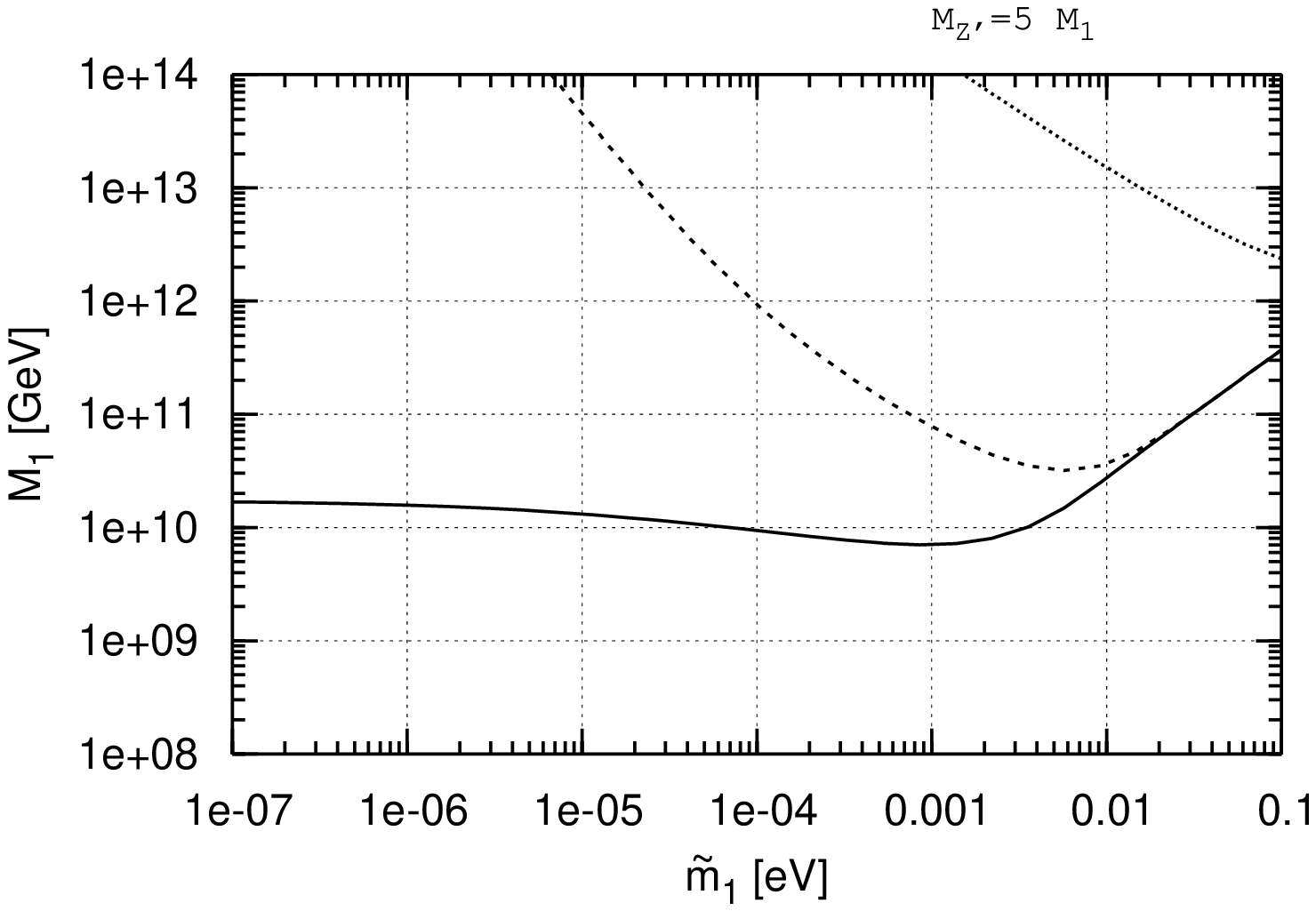 ,width=0.5\textheight,angle=270}}}
\caption[]{The allowed regions in the $\tilde m_1 - M_1$ plane for $M_{Z'}=
  5 M_1$ and different values of the reheating temperature: the regions
  above the solid, dashed and dotted curves
  correspond respectively to $T_{rh} = M_1/3, M_1/5$ and $0.1 M_1$.} 
\label{fig:4MZ5}
}  

In the two cases ($M_{Z'}=5, 100 M_1$) illustrated in figs.~\ref{fig:4MZ100}
and \ref{fig:4MZ5} the lowest values allowed for $T_{rh}$ are quite above $6
\times 10^8$~GeV, which is the lowest possible value of $T_{rh}$ for
hierarchical leptogenesis scenarios where the heavy neutrinos are thermally
produced. That bound  corresponds to the idealized situation in which the
main interaction that produces the heavy neutrinos is very fast before decoupling abruptly at a
certain value of 
$z$ (say $z=z_{fo}$). This is because when that kind of interaction is present
an equilibrium population of $N_1$ can be achieved for a reheating temperature
 as low as $M_1/z_{fo}$, while for $z > z_{fo}$ the (CP conserving) interaction
 effectively vanishes and hence all the neutrinos disappear
 via the CP violating Yukawa couplings. In this case the final asymmetry
 (in the weak washout regime) would be 
 given by $Y_{B-L}^f \approx - \epsilon Y_N^{relic} = - \epsilon
 Y_N^{eq}(z=z_{fo})$ and taking into account that the maximum CP asymmetry is
 proportional to $M_1$ it's straightforward to find that the optimum values for
 $z_{fo}$ are near unity\footnote{\samepage{In fact, the optimum situation for obtaining low reheating
  temperatures happens when the interaction decouples abruptly at $z$ values
  somewhat larger than unity, while for $z_{fo} < 0.5$ or $z_{fo} > 5$ the lower
  bound on $T_{rh}$ is greater than the ideal bound by a factor of 2 or more,
  even for the idealized type of interactions just described.}}. However, in a realistic model the
 interactions depart from equilibrium gradually, so the bound on $T_{rh}$ is
 always above the ideal one. To check this we display in fig.~\ref{fig:5} for the whole
 range of masses $M_{Z'}/M_1 < 100$ (i.e. for case (iii)) the lower
 bounds on $M_1$ and $T_{rh}$ for the weak
 washout regime (we have taken $\tilde m_1 = 10^{-6}$~eV but the results are
 almost  the same for any value of $\tilde m_1 \ll 10^{-3}$~eV). Several things are
 apparent from the plot. First we see that the lowest possible value of $T_{rh}$ in the $Z'$
 model we are studying is
 approximately equal to $1.2 \times 10^{9}$~GeV, i.e. a factor of two greater
 than the ideal bound. This value of $T_{rh}$ is possible only for a special
 range of $Z'$ masses around $M_{Z'} = 20 M_1$, for which the corresponding
 $U(1)_{Y'}$ gauge interactions depart from equilibrium at $z \sim 1$ (note
 that in this range the bounds on $M_1$ and $T_{rh}$ are very similar), while
 for $M_{Z'}$ approaching $100 M_1$ the bound is 5 times greater than the
 ideal one. On the other hand, the bound
 on $M_1$ shows the  behavior already explained in the previous
 section: when $M_{Z'}$ is small\footnote{\samepage{We have included $Z'$ masses as low
   as $M_1$ in fig.~\ref{fig:5} with the purpose of testing different
   situations (i.e. interactions which produce $N_1$ and have different
   decouple behavior), but we remind that the correct calculation of the bounds for
   very low $M_{Z'}$ must take into account the effects of the $\chi$ field
   (it is to expect that it's inclusion will suppress even more the efficiency
 but won't change the picture
   qualitatively).}} the suppression of the efficiency must be
 compensated with large values of the CP asymmetry (and hence of $M_1$), while
 for $M_{Z'}$
 approaching $100 M_1$ the interactions mediated by $Z'$ depart from
 equilibrium when the $N_1$ are still relativistic, so the bound on $M_1$
 reaches it's lowest possible value (equal to $6
\times 10^8$~GeV ). Note also that the lowest efficiencies occur when $M_{Z'}
\approx 2 M_1$, as has been explained before. 
\FIGURE{
\centerline{\protect\hbox{
\epsfig{file=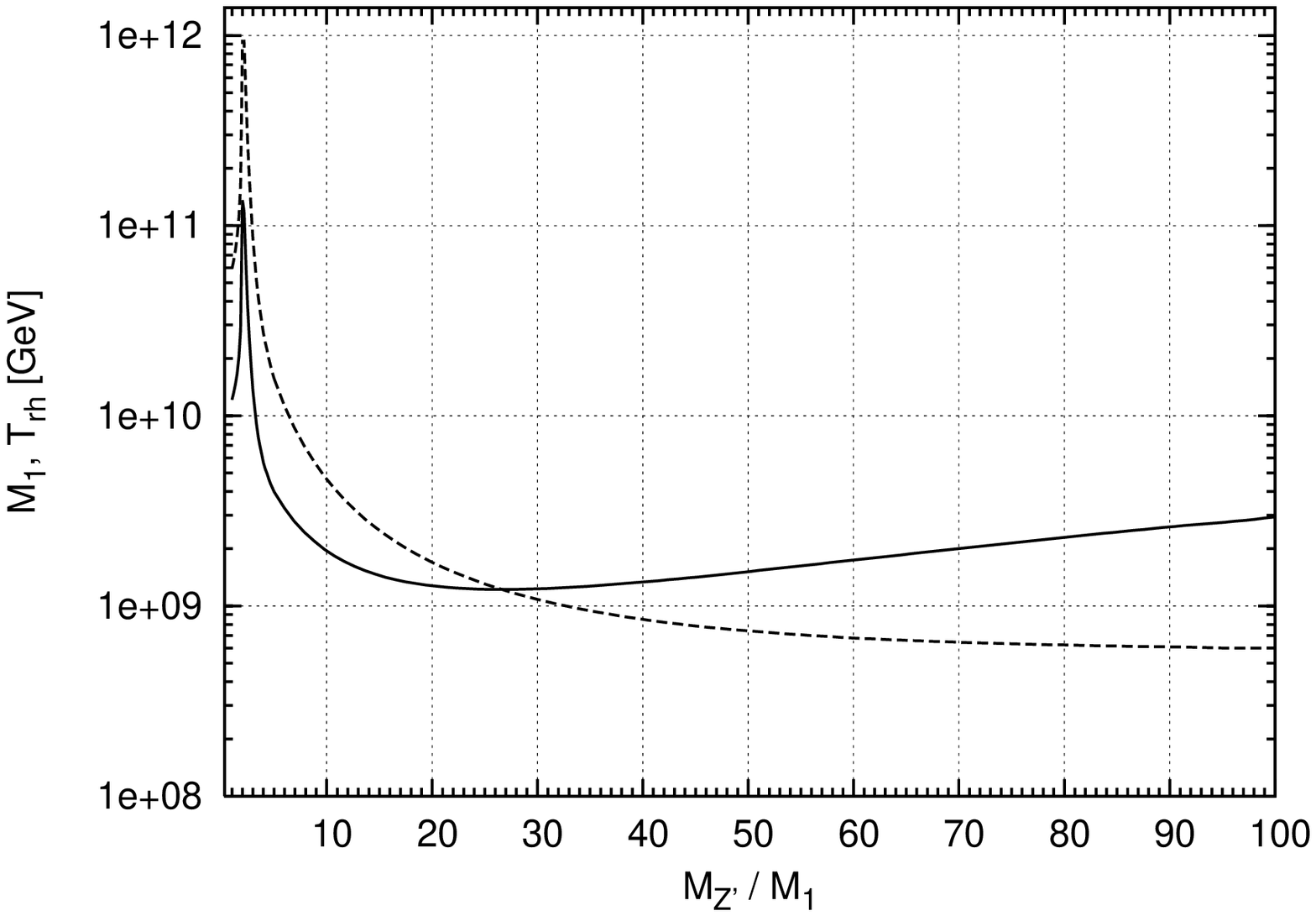 ,width=0.45\textheight,angle=270}}}
\caption[]{The lower bound on $M_1$ (dashed curved) and on $T_{rh}$ (solid
  curve) as a function of $M_{Z'}/M_1$ for $\tilde m_1=10^{-6}$~eV.} 
\label{fig:5}
}  

Finally, it's also clear from fig.~\ref{fig:5} that for
low $Z'$ masses the bound on $T_{rh}$ can be several (up to 7) times smaller than the
corresponding bound on $M_1$, as discussed at the beginning of this
section. However, this works only for comparatively
large $M_1$ values and this is also true in the strong washout regime (see
fig.~\ref{fig:4MZ5} for the case $M_{Z'}= 5M_1$), so the bound for $T_{rh}$ is
several times greater than the ideal one also for these cases. We conclude that, in hierarchical scenarios where the heavy neutrinos are thermally produced, even in the presence of additional $Z'$ gauge bosons the reheating temperature needs to be larger than $10^9$~GeV and hence this cannot be of much help in
relation to the gravitino problem affecting some supersymmetric scenarios.
\section{Conclusions}
\label{sec:concl}
The existence of neutral gauge bosons coupled to the heavy neutrinos notably
affects the leptogenesis picture in the weak washout regime. The main new
ingredient with respect to the simplest thermal leptogenesis scenarios is that
they allow the production and destruction of the heavy neutrinos without
generating a CP asymmetry. When the $Z'$ bosons are not very heavy compared to
the reheating temperature ($M_{Z'} \lesssim 10 T_{rh}$) an equilibrium population
of $N_1$ is always achieved before the neutrinos become non-relativistic (as
long as $T_{rh} > M_1$). Moreover, if the new gauge bosons are not too light
($M_{Z'} \gtrsim 100 M_1$) the corresponding gauge interactions depart from
equilibrium before the heavy neutrinos become non-relativistic and in this case
the
efficiency reaches its maximum possible value.  On the other hand, for lighter
$Z'$ bosons ($M_{Z'} \lesssim 100 M_1$) the gauge interactions remain in equilibrium until a temperature which is
smaller than  $M_1$ and hence the $N_1$
can partially disappear without producing a lepton asymmetry (via the interactions
$\proname{N_1 N_1}{\ell^+ \ell^-, h^+ h^-}$). The suppression effect induced by
these interactions is greatest for $M_{Z'} \approx 2 M_1$.

We have also shown that the presence of $Z'$ bosons not much heavier than
$N_1$ allows to have reheating temperatures a few times (even up to a factor seven) smaller than $M_1$
still obtaining large efficiencies (compared to the cases when the heavy
neutrinos are produced only via the Yukawa interactions). However, for
hierarchical leptogenesis scenarios in which the heavy neutrinos are produced
thermally, the minimum
reheating temperature required for successful leptogenesis is always quite
above the lowest possible value for $M_1$ (equal to $6
\times 10^8$~GeV for a cosmic baryon asymmetry equal to $Y_B=8.7 \times
10^{-11}$). In the $Z'$ model we have analyzed the lowest bound on $T_{rh}$ is
two times that value, but this happens only for very special values of $M_{Z'}$
(around $20 M_1$), while for other values of the $Z'$ mass the required value of the reheating temperature increases.   

\section*{Acknowledgments}
The work of J.~R.~is supported by research grants FPA2007-66665 and
2005SGR00564. It is also supported by the Consolider-Ingenio 2010 Program CPAN
(CSD2007-00042). The work of E.~R.~is partially supported by the grant PICT 13562 of the ANPCyT.
\bibliographystyle{JHEP}
\bibliography{referencias_leptogenesisV2}

\providecommand{\href}[2]{#2}\begingroup\raggedright\begin{thebibliography}{10}

\bibitem{fukugita86}
M.~Fukugita and T.~Yanagida, {\it {Baryogenesis Without Grand Unification}},
  {\em Phys. Lett.} {\bf B174} (1986) 45.

\bibitem{davidson02}
S.~Davidson and A.~Ibarra, {\it {A lower bound on the right-handed neutrino
  mass from leptogenesis}},  {\em Phys. Lett.} {\bf B535} (2002) 25,
  [\href{http://xxx.lanl.gov/abs/hep-ph/0202239}{{\tt hep-ph/0202239}}].

\bibitem{komatsu08}
{E. Komatsu {\it et al.} [WMAP Collaboration]}, {\it {Five-Year Wilkinson
  Microwave Anisotropy Probe (WMAP) Observations: Cosmological
  Interpretation}},  \href{http://xxx.lanl.gov/abs/astro-ph/0803.0547}{{\tt
  astro-ph/0803.0547}}.

\bibitem{hambye03}
T.~Hambye, Y.~Lin, A.~Notari, M.~Papucci, and A.~Strumia, {\it {Constraints on
  neutrino masses from leptogenesis models}},  {\em Nucl. Phys.} {\bf B695}
  (2004) 169, [\href{http://xxx.lanl.gov/abs/hep-ph/0312203}{{\tt
  hep-ph/0312203}}].

\bibitem{abada06II}
A.~Abada, S.~Davidson, A.~Ibarra, F.~Josse-Michaux, M.~Losada, and A.~Riotto,
  {\it {Flavour matters in leptogenesis}},  {\em JHEP} {\bf 09} (2006) 010,
  [\href{http://xxx.lanl.gov/abs/hep-ph/0605281}{{\tt hep-ph/0605281}}].

\bibitem{nardi07II}
E.~Nardi, J.~Racker, and E.~Roulet, {\it {CP violation in scatterings, three
  body processes and the Boltzmann equations for leptogenesis}},  {\em JHEP}
  {\bf 09} (2007) 090, [\href{http://xxx.lanl.gov/abs/hep-ph/0707.0378}{{\tt
  hep-ph/0707.0378}}].

\bibitem{plumacher96}
M.~Pl{\"u}macher, {\it {Baryogenesis and lepton number violation}},  {\em Z.
  Phys.} {\bf C74} (1997) 549,
  [\href{http://xxx.lanl.gov/abs/hep-ph/9604229}{{\tt hep-ph/9604229}}].

\bibitem{cosme04}
N.~Cosme, {\it {Leptogenesis, neutrino masses and gauge unification}},  {\em
  JHEP} {\bf 08} (2004) 027,
  [\href{http://xxx.lanl.gov/abs/hep-ph/0403209}{{\tt hep-ph/0403209}}].

\bibitem{abada06}
A.~Abada, S.~Davidson, F.~X. Josse-Michaux, M.~Losada, and A.~Riotto, {\it
  {Flavour issues in leptogenesis}},  {\em JCAP} {\bf 0604} (2006) 004,
  [\href{http://xxx.lanl.gov/abs/hep-ph/0601083}{{\tt hep-ph/0601083}}].

\bibitem{nardi06}
E.~Nardi, Y.~Nir, E.~Roulet, and J.~Racker, {\it {The importance of flavor in
  leptogenesis}},  {\em JHEP} {\bf 01} (2006) 164,
  [\href{http://xxx.lanl.gov/abs/hep-ph/0601084}{{\tt hep-ph/0601084}}].

\bibitem{buchmuller01}
W.~Buchm{\"u}ller and M.~Pl{\"u}macher, {\it {Spectator processes and
  baryogenesis}},  {\em Phys. Lett.} {\bf B511} (2001) 74,
  [\href{http://xxx.lanl.gov/abs/hep-ph/0104189}{{\tt hep-ph/0104189}}].

\bibitem{nardi05}
E.~Nardi, Y.~Nir, J.~Racker, and E.~Roulet, {\it {On Higgs and sphaleron
  effects during the leptogenesis era}},  {\em JHEP} {\bf 01} (2006) 068,
  [\href{http://xxx.lanl.gov/abs/hep-ph/0512052}{{\tt hep-ph/0512052}}].

\bibitem{giudice04}
G.~F. Giudice, A.~$\text{Notari}$, M.~Raidal, A.~Riotto, and A.~Strumia, {\it
  {Towards a complete theory of thermal leptogenesis in the SM and MSSM}},
  {\em Nucl. Phys.} {\bf B685} (2004) 89,
  [\href{http://xxx.lanl.gov/abs/hep-ph/0310123}{{\tt hep-ph/0310123}}].

\bibitem{carlier99}
S.~Carlier, J.~M. Fr\`ere, and F.~S. Ling, {\it {Gauge dilution and
  leptogenesis}},  {\em Phys. Rev.} {\bf D60} (1999) 096003,
  [\href{http://xxx.lanl.gov/abs/hep-ph/9903300}{{\tt hep-ph/9903300}}].

\bibitem{frere08}
J.~M. Fr\`ere, T.~Hambye, and G.~Vertongen, {\it {Is leptogenesis falsifiable
  at LHC?}},  \href{http://xxx.lanl.gov/abs/hep-ph/0806.0841}{{\tt
  hep-ph/0806.0841}}.

\bibitem{buchmuller04}
W.~Buchm{\"u}ller, P.~Di~Bari, and M.~Pl{\"u}macher, {\it {Leptogenesis for
  pedestrians}},  {\em Ann. Phys.} {\bf 315} (2005) 305,
  [\href{http://xxx.lanl.gov/abs/hep-ph/0401240}{{\tt hep-ph/0401240}}].

\end{thebibliography}\endgroup
\end{document}